\documentclass[10pt,twocolumn,letterpaper]{article}

\usepackage[pagenumbers]{wacv} %
\usepackage{graphicx}
\usepackage{amsmath}
\usepackage{amssymb}
\usepackage{booktabs}

\usepackage[pagebackref,breaklinks,colorlinks]{hyperref}
\usepackage[accsupp]{axessibility} %

\usepackage[capitalize]{cleveref}
\crefname{section}{Sec.}{Secs.}
\Crefname{section}{Section}{Sections}
\Crefname{table}{Table}{Tables}
\crefname{table}{Tab.}{Tabs.}

\def\wacvPaperID{2120} %
\def\confName{WACV}
\def\confYear{2025}

\graphicspath{{figs/}}

\begin{document}

\title{Enhancing predictive imaging biomarker discovery through treatment effect analysis}

\author{Shuhan Xiao$^{1,2}$\qquad
Lukas Klein$^{3,4,5}$\qquad
Jens Petersen$^{1}$\qquad
Philipp Vollmuth$^{1,6,7}$,\\
Paul F. Jaeger$^{3,5}$\thanks{These authors contributed equally to this work.}\qquad
Klaus H. Maier-Hein$^{1,2,5,8}$\footnotemark[1]\\
\small $^{1}$German Cancer Research Center (DKFZ) Heidelberg, Division of Medical Image Computing, Germany\quad
$^{2}$Faculty of Mathematics\\\small and Computer Science, Heidelberg University, Germany\quad
$^{3}$DKFZ Heidelberg, Interactive Machine Learning Group, Germany\\\small
$^{4}$Institute for Machine Learning, ETH Zürich, Switzerland\quad
$^{5}$DKFZ Heidelberg, Helmholtz Imaging, Germany\\\small 
$^{6}$Division for Computational Radiology Clinical AI (CCIBonn.ai), Clinic for Neuroradiology,\\\small University Hospital Bonn, Germany\quad
$^{7}$Medical Faculty Bonn, University of Bonn, Germany\\\small
$^{8}$Pattern Analysis and Learning Group, Department of Radiation Oncology, Heidelberg University Hospital, Germany\\
{\tt\small s.xiao@dkfz-heidelberg.de}
}

\maketitle

\begin{abstract}
Identifying predictive covariates, which forecast individual treatment effectiveness, is crucial for decision-making across different disciplines such as personalized medicine. These covariates, referred to as biomarkers, are extracted from pre-treatment data, often within randomized controlled trials, and should be distinguished from prognostic biomarkers, which are independent of treatment assignment. 
Our study focuses on discovering predictive imaging biomarkers, specific image features, by leveraging pre-treatment images to uncover new causal relationships. 
Unlike labor-intensive approaches relying on handcrafted features prone to bias, we present a novel task of directly learning predictive features from images.
We propose an evaluation protocol to assess a model's ability to identify predictive imaging biomarkers and differentiate them from purely prognostic ones by employing statistical testing and a comprehensive analysis of image feature attribution.
We explore the suitability of deep learning models originally developed for estimating the conditional average treatment effect (CATE) for this task, which have been assessed primarily for their precision of CATE estimation while overlooking the evaluation of imaging biomarker discovery.
Our proof-of-concept analysis demonstrates the feasibility and potential of our approach in discovering and validating predictive imaging biomarkers from synthetic outcomes and real-world image datasets. 
Our code is available at \url{https://github.com/MIC-DKFZ/predictive_image_biomarker_analysis}.
\end{abstract}

\section{Introduction}
\begin{figure}[t]
  \centering
   \includegraphics[trim=0cm 22.5cm 43cm 0cm, clip, width=0.75\linewidth]{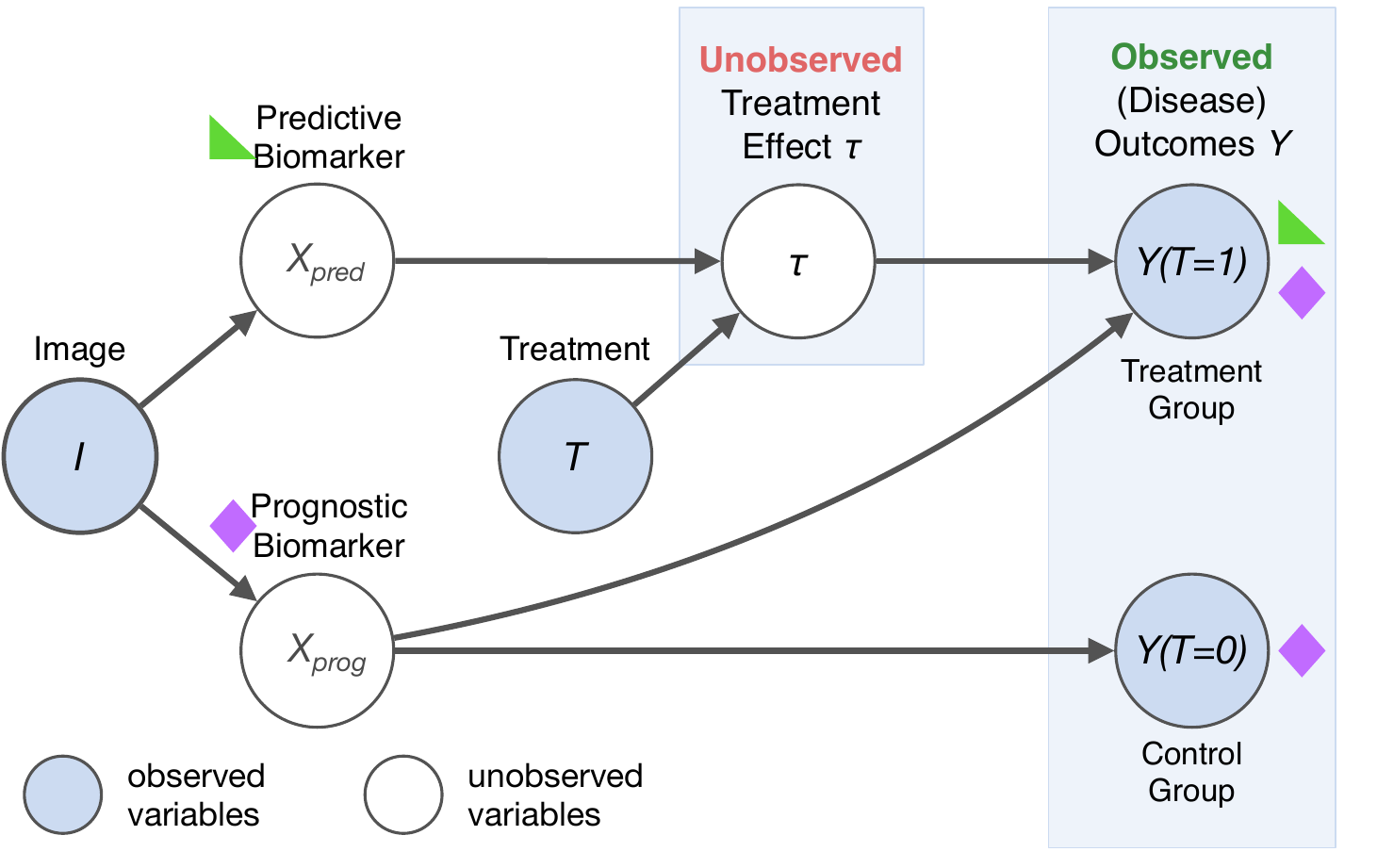}
   \caption{Relationship between biomarkers $x_{\mathit{prog}}$ and $x_{\mathit{pred}}$, outcomes $Y(T)$ depending on the treatment $T$ and the treatment effect $\tau$. Since both potential outcomes $Y_i(T=0)$ and $Y_i(T=1)$ cannot be observed for the same individual simultaneously it is impossible to infer the individual treatment effect directly.} %
   \label{fig:diagram}
\end{figure}
Identifying predictive biomarkers is crucial for determining which subgroup of individuals will have a positive treatment effect and ultimately for making informed treatment decisions across different fields such as medical treatments, environmental strategies, and economic policies. Precision medicine, for example, relies on predictive biomarkers to tailor interventions to individual patients and ensure optimized patient outcomes.
Generally, a biomarker is a measurable characteristic associated with an individual's outcome such as disease progression or physiologic measures~\cite{lohr1988outcome}. Although the term originally stems from the biomedical field, we use it more broadly in this paper to refer to features or covariates in general contexts. A biomarker is predictive when it acts as a driver of treatment effect heterogeneity~\cite{curth2021really}. Predictive biomarkers are treatment-specific, while prognostic biomarkers are associated with the outcome independent of treatment assignment~\cite{ballman2015biomarker}, as illustrated in Fig.~\ref{fig:diagram}.
The discovery of predictive biomarkers is key to not only explaining the causal mechanisms behind treatment effects and supporting informed treatment decisions but also to driving the development of novel treatments.
In particular, there has been a growing interest in leveraging the vast amount of non-invasively acquired information provided by different imaging modalities to discover so-called imaging biomarkers, especially predictive imaging biomarkers~\cite {oconnor_imaging_2017}.

In previous research, the discovery process of predictive imaging biomarkers involves handcrafted radiomics features (\eg shape, intensity, and texture of tumors or lesions~\cite{limkin2017promises,lambin2017radiomics,kumar2012radiomics,bortolotto2021radiomics}) as candidates to determine their predictive performance. This process typically contains several steps including segmentation to define regions of interest, feature extraction, and feature selection.

While machine learning approaches have been employed to facilitate the discovery of imaging biomarkers~\cite{kickingereder2016large,chaddad2019radiomics,parmar2015machine,park2018radiomics,limkin2017promises,lambin2017radiomics,lou2019image}, the training processes rely on handcrafted radiomics-based features and have the risk of introducing human bias, as shown in \cite{hosny2019handcrafted}. %
Some approaches directly aim at discovering predictive biomarkers and distinguishing them from prognostic ones~\cite{sechidis2018distinguishing,boileau2023flexible,zhu2023identification,bahamyirou2022doubly}, but are limited to tabular input data. More flexibility and adaptability are offered by deep learning (DL)-based conditional average treatment effect (CATE) estimation methods~\cite{shalit2017estimating, alaa2017deep,hassanpour2019learning,shi2019adapting,curth2021nonparametric,zhang2021treatment}, which have the potential to identify predictive biomarker candidates from a set of tabular covariates as well~\cite{crabbe2022benchmarking,boileau2023flexible}.
CATE estimation differs from a standard supervised learning task and requires different modeling approaches as the ground truth for our quantity of interest -- the individual treatment effect -- is not available. This is due to the fundamental problem of causal inference~\cite{holland1986statistics}: It is impossible to observe both potential outcomes, treated and untreated, from the same individual simultaneously, yet they are necessary to compute the individual treatment effect.  
For CATE estimation, the presence of strong prognostic biomarkers, which is frequently encountered in practice, can negatively impact the performance of CATE estimators, even though they are not relevant for the treatment effect and, thus, treatment decision-making. 
For instance, CATE estimators can mistakenly identify prognostic as predictive biomarkers, as studies have shown~\cite{hermansson2021discovering,crabbe2022benchmarking,sechidis2018distinguishing}, which may lead to ineffective or even harmful treatment recommendations. It is therefore essential to ensure that these methods can distinguish the two types of biomarkers.

CATE estimation methods have been originally designed for tabular inputs and remain a widely unexplored topic in the context of image inputs. In response to this gap, recent advancements have adapted DL-based CATE estimation methods to estimate treatment effects not only from medical images~\cite{durso2022personalized,durso-finley_improving_2023,ma2023treatment,jiang2024deep} but also other types of images~\cite{jerzak2022image,takeuchi2021grab,jiang2023estimating}.
Yet, none of these image-based methods directly describe how predictive biomarkers can be identified and interpreted or address how well models manage to do so, which is an important but often overlooked performance metric to consider when evaluating CATE estimation methods, as noted in~\cite{curth2021really}. To conduct such an evaluation, a benchmarking environment was proposed in \cite{crabbe2022benchmarking}, albeit only applicable to tabular data.

Adapting the evaluation of predictive biomarker discovery from tabular data to images introduces a significant challenge: Extracting imaging biomarkers is complicated by the high-dimensional and structured nature of image data, which lacks distinct, pre-defined features. Consequently, a critical step in interpreting these biomarkers is determining the specific image features on which a black-box CATE estimation model depends. This step is also vital for drug development and clinical decision-making.

\begin{figure*}[!t]
  \centering
   \includegraphics[trim=0cm 12cm 16.8cm 0cm, clip, width=0.93\linewidth]{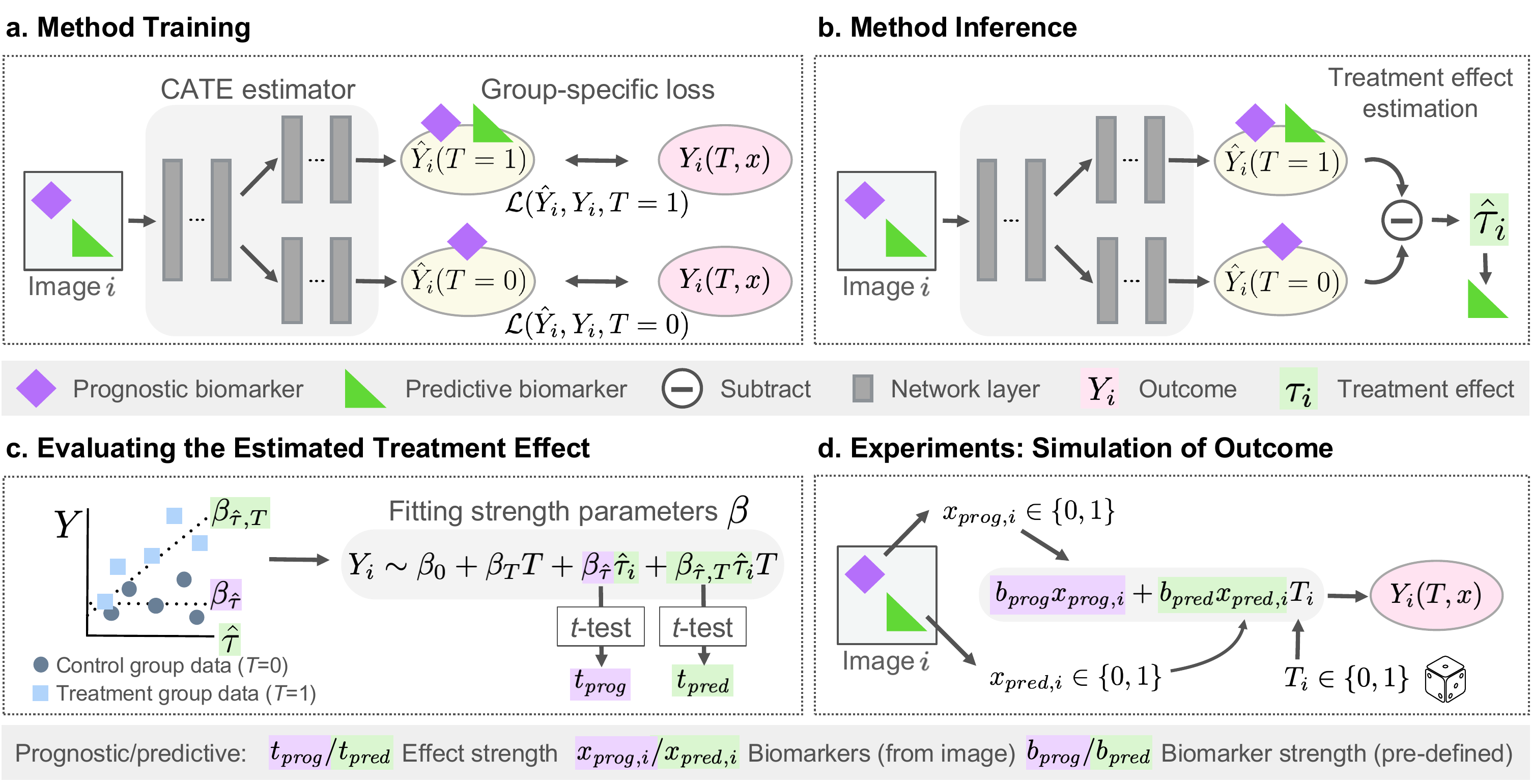}
   \caption{Overview of the identification of predictive biomarkers from pre-treatment images. The (\textbf{a}) training and (\textbf{b}) inference step employs a two-headed architecture to estimate treatment effects $\hat{\tau}$ from images. In the evaluation step (\textbf{c}) the predictive strength of the estimated $\hat{\tau}$, the predictive biomarker candidate, is assessed using regression. In our simulation experiments (\textbf{d}), the outcome data $Y_i$ used in our experiments are simulated with image features from ground truth annotations and randomly assigned treatments $T$.} %
   \label{fig:overview}
\end{figure*}

In this paper, we define a novel task in response to the challenges above: discovering predictive imaging biomarkers directly from image data in a data-driven way, without requiring handcrafted features or a separate feature extraction step. We introduce a new evaluation protocol tailored to this task and demonstrate as a proof-of-concept how a DL-based CATE estimation model can be applied in practice (Fig.~\ref{fig:overview}). 
Our evaluation protocol includes two components: (1) statistical testing to investigate the estimated predictive biomarker strength, and (2) explainable artificial intelligence (XAI) methods~\cite{simonyan2013deep,sundararajan_axiomatic_2017,erion_improving_2020,Selvaraju2017,Springenberg2014} to enable the verification and interpretation of the discovered predictive imaging biomarker candidates. 

We also propose and conduct experiments to validate our evaluation protocol on real image data using pre-defined imaging biomarkers with varying strengths of predictive and prognostic effects on synthetic outcomes. This setup, for benchmarking and model development, enables assessing a model's ability to identify and interpret predictive imaging biomarkers. Experiments on natural and medical images highlight the potential of an image-based CATE estimator to address our task, showcasing the model's capability to identify predictive imaging biomarkers with greater predictive strength compared to a baseline that does not distinguish between prognostic and predictive effects.

\section{Methods}

\subsection{Treatment heterogeneity and predictive biomarkers}
We describe how treatment effects, which cannot be observed directly, can be estimated from data by introducing the concept of potential outcomes. Here, we consider pre-treatment images and data collected through randomized controlled trials (RCTs), the typical experimental setting for discovering biomarkers. The relation between outcomes, defined by a problem-specific measure of interest, and treatment effect has been described by the Neyman-Rubin causal model~\cite{rubin2005causal}, where the individual treatment effect (ITE) for an individual $i$ is defined as the difference between potential outcomes $Y_i(T)$, $\mathrm{ITE} := Y_i(T=1) - Y_i(T=0)$. Here, we assume a binary treatment variable $T\in \{ 0,1 \}$ for whether a treatment is applied or not. In RCTs, $T$ is randomly assigned and indicates whether an individual belongs to the control group ($T=0$) or treatment group ($T=1$). Since it is not possible to observe the counterfactual outcomes and thus measure the ITE due to the fundamental problem of causal inference, in practice, the conditional average treatment effect (CATE) $\tau$
\begin{equation}
\begin{split}
    \tau(x) &:= \mathbb{E} \left[ Y(T=1) - Y(T=0) | X=x \right]\label{eq:cate}
\end{split}
\end{equation}
is estimated instead. The CATE depends on observable pre-treatment covariates $x\in X$, which can for example be extracted from images $I$. While such covariates that measure image features are often called imaging biomarkers in biomedical applications, we use ``imaging biomarkers'' as a more general term. %
Only heterogeneous treatment effects, \ie effects that vary among individuals and covariates $x$, are relevant for making treatment decisions or subgroup selection. Therefore, we are interested in identifying covariates that directly contribute towards the heterogeneous treatment effect and interact with the treatment, also known as predictive biomarkers. Under the common assumption that prognostic effects $f_{\mathit{prog}}$ and predictive effects $f_{\mathit{pred}}$ are additive~\cite{sechidis2018distinguishing, kunzel2019metalearners,curth2021really, hermansson2021discovering} as in
\begin{equation}
    \mathbb{E} \left[ Y(x) \right] = f_{\mathit{prog}}(x) + f_{\mathit{pred}}(x)T,
    \label{eq:additiveeffects}
\end{equation}
the CATE defined in Eq.~(\ref{eq:cate}) yields $f_{\mathit{pred}}(x)$, which only depends on predictive biomarkers $x_{\mathit{pred}}$.
In this case, treatment effect estimation automatically separates prognostic and predictive effects and thus identifies predictive biomarkers $x_{\mathit{pred}}$. Generally, a biomarker can be both prognostic and predictive at the same time if it contributes to both $f_{\mathit{prog}}(x)$ and $f_{\mathit{pred}}(x)$.
Figure \ref{fig:diagram} depicts the relationship between biomarkers $x_{\mathit{pred}}$ or $x_{\mathit{prog}}$ and outcomes $Y$.

\subsection{Image-based treatment effect estimator}
To enable the discovery of predictive imaging biomarkers, we leverage neural network-based CATE estimators adapted for image inputs. For our experiments, we modify a TARNet model~\cite{shalit2017estimating}, originally designed for tabular inputs, similar to the adaptation described in~\cite{durso2022personalized}. The network has shared convolutional layers as encoders for learning the similarities between the control and treatment groups arising from prognostic effects~\cite{curth2021inductive}, and two treatment-specific heads for predicting the outcomes $Y(T)$. 
During the training (Fig.~\ref{fig:overview}a), we apply the loss to the corresponding head, depending on which RCT group the input data belongs to. In each training step, the total loss is the sum of the loss of the control group head output and the treatment group head output, so that the weights of both heads are updated. 
During inference (see Fig.~\ref{fig:overview}b), the CATE is estimated by subtracting the model's control group output from the treatment group output: $\hat{\tau}= \hat{Y}_i(T=1) - \hat{Y}_i(T=0)$.

In contrast to the two-headed model, we expect a single-head model to learn to predict the average outcome across groups from both predictive and prognostic biomarkers and not differentiate between the treatment group or control group. The predicted outcome of such a network is the composition of both predictive and prognostic effects. It is used as a baseline to validate whether the CATE estimator could successfully discover a predictive biomarker. Implementation details are described in the Supplementary section \textbf{A.4}.

\subsection{Proposed evaluation protocol} \label{sec:eval}
\subsubsection{Statistical evaluation of the predictive strength}
To verify whether the model has identified a predictive effect -- that is, whether the estimated CATE $\hat{\tau}$ is indeed predictive and can be considered a predictive biomarker candidate -- we test the interaction between biomarker candidate and treatment, as seen in Fig.~\ref{fig:overview}c. Such an evaluation is also performed in clinical practice~\cite{ballman2015biomarker,polley2013statistical}. %
We assume a linear relationship between biomarkers and outcome (Eq.~(\ref{eq:additiveeffects})) and perform a linear regression of the outcomes $Y$ using
\begin{equation}
    \beta_0 + \beta_T \, T + \beta_{\hat{\tau}} \, {\hat{\tau}} + \beta_{{\hat{\tau}},T} \, {\hat{\tau}} \, T \sim Y,
    \label{eq:regression}
\end{equation}
which includes an interaction term $\beta_{{\hat{\tau}},T} \, {\hat{\tau}}$ and coefficients $\beta_i$. We test the null hypothesis that the biomarker-treatment interaction coefficient is $\beta_{\hat{\tau},T}=0$ using the Student's $t$-test with the $t$-value $t_{\beta_{\hat{\tau},T}}$ test statistic, which is proportional to the estimated $\hat{\beta}_{\hat{\tau},T}$. This test is additionally repeated with the other fit coefficients $\beta_i$. %
The $t$-value ratio $t_{\beta_{\hat{\tau},T}}/t_{\beta_{\hat{\tau}}}=:t_{\mathit{pred}}/t_{\mathit{prog}}$ can be used as an indicator for the predictive strength of the estimated CATE $\hat{\tau}$ compared to its prognostic strength.
To estimate the experimental lower (indicating a prognostic biomarker) and upper (indicating a predictive biomarker) bound for the relative predictive strength, we conduct the same evaluation, replacing $\hat{\tau}$ in Eq.~(\ref{eq:regression}) with either the purely prognostic or a purely predictive ground truth biomarker $x_{\mathit{prog},\mathit{pred}}$. 

\subsubsection{Interpretation using feature attribution methods} %
We also investigate which input image features the trained model is sensitive to when predicting the CATE $\hat{\tau}$ and whether they correspond to predictive imaging biomarkers. Since a direct quantitative assessment is not straightforward for general image features, unlike for tabular data, we rely on visual explanations through attribution maps~\cite{simonyan2013deep} instead.
To this end, we employ the XAI methods expected gradients (EG)~\cite{erion_improving_2020} and guided gradient-weighted class activation mapping (GGCAM)~\cite{Selvaraju2017, Springenberg2014} to generate attribution maps from the trained model and input images. The attribution maps enable us to visually analyze how much individual pixels contribute to either the prognostic effect via the attribution map of the control group head prediction $\hat{Y}(T=0)$ or the predictive effect via the attribution map of the estimated CATE $\hat{Y}(T=1)-\hat{Y}(T=0)$.  %

\subsection{Simulation of imaging biomarkers and outcomes for validation}
To study the CATE estimator's ability to identify predictive imaging biomarkers in the presence of prognostic ones, we conduct experiments on data with varying predictive and prognostic biomarker strengths.
Since ground truth counterfactual treatment outcomes are unavailable in real data, we generate synthetic data to experimentally verify the model and simulate the ground truth treatment outcomes (Fig.~\ref{fig:overview}d). 
Our proposed approach simulates outcomes based on imaging biomarkers by assigning image features to biomarker values $x_{\mathit{prog},\mathit{pred}}$ instead of simulating outcomes directly from tabular biomarkers. This entails selecting features from available image information such as attributes, class labels, or radiomics features, as shown in Fig.~\ref{fig:biomarkers}. In our examples, the biomarkers are either purely prognostic or predictive and may be binary or continuous depending on the dataset.
The outcomes $Y$ are then generated according to a simple linear function:
\begin{equation}
    Y(T,x) = b_{\mathit{prog}} x_{\mathit{prog}} + b_{\mathit{pred}} \, x_{\mathit{pred}} T, 
    \label{eq:biomarker}
\end{equation} 
assuming no offset $b_0$ and constant treatment effect $b_{T}$ for simplicity, similar to a case considered in~\cite{krzykalla2020exploratory}.
An important aspect of using simulated outcomes is that we can control the size of prognostic or predictive effects by adjusting the parameters $b_{\mathit{prog},\mathit{pred}}$. The biomarker parameter strength ratio $b_{\mathit{pred}}/b_{\mathit{prog}}$ can be interpreted as a measure of the signal-to-noise ratio of the predictive effect in the input data.
Here, in an RCT setting, the treatment variable $T\in\{0,1\}$ is assigned with probabilities $p(T)=0.5$.

\subsection{Experimental Setup}
\subsubsection{Datasets and imaging biomarker features} \label{sec:datasets}
\begin{figure}[ht]
  \centering
   \includegraphics[trim=0cm 22cm 18cm 0cm, clip, width=1.0\linewidth]{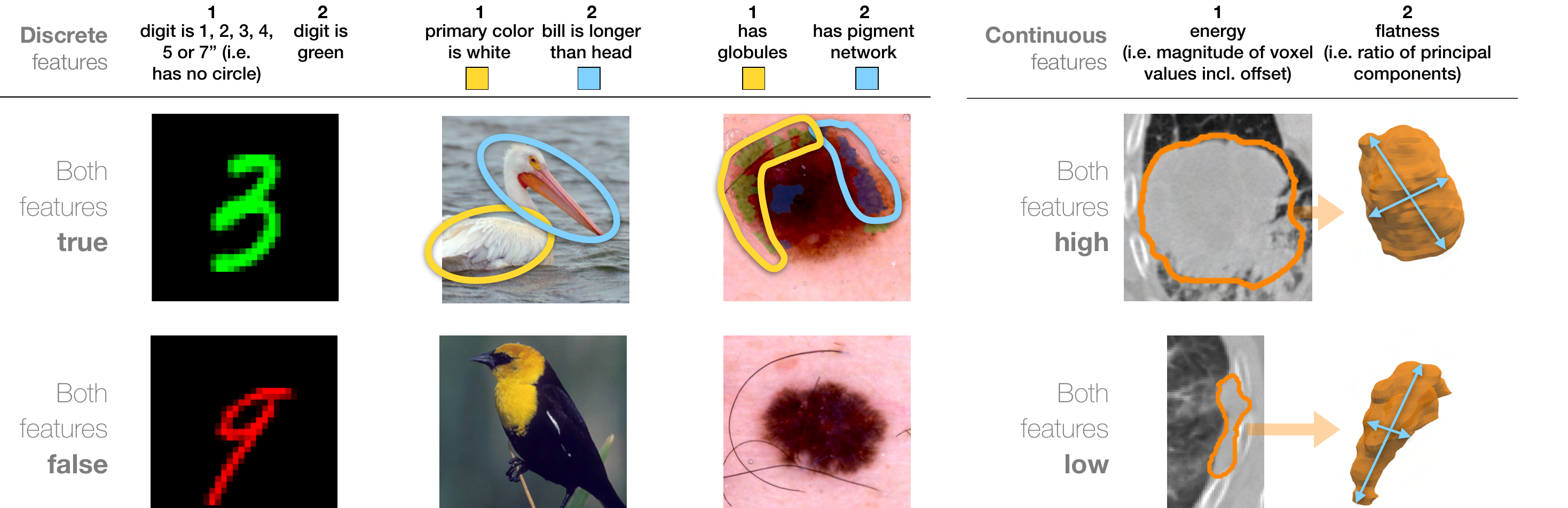}
   \caption{Image features from the four datasets, where either feature 1 or 2 is designated as predictive or prognostic biomarkers. ISIC 2018 skin lesion features are shown with ground truth masks. Globules (light green mask) manifest as darker dots, pigment networks have dark grid-like patterns of streaks with lighter ``holes'' (dark blue mask). The NSCLC-Radiomics images display tumor segmentation outlines of a 2D slice (left) or corresponding 3D volumes (right). Examples on the bottom row depict images where both biomarkers are either absent or have a low value.} %
   \label{fig:biomarkers}
\end{figure}

We evaluate our CATE estimator on four diverse publicly available datasets also shown in Fig.~\ref{fig:biomarkers}: colored digits (MNIST~\cite{deng2012mnist,arjovsky2019invariant}) with semi-synthetic image features, images of birds (CUB-200-2011~\cite{WahCUB_200_2011}) as an example of a natural image dataset, as well as skin lesion images (ISIC 2018~\cite{codella2019skin, tschandl2018ham10000}) and 3D lung computed tomography (CT) scans of non-small cell lung cancer (NSCLC) tumors (NSCLC-Radiomics~\cite{aerts2014}) as real-world medical datasets.

\textbf{Colored MNIST (CMNIST).} We adapt the MNIST dataset and introduce color as an image feature. The color of the digits is determined based on the random variable $x_i$ sampled from a binomial distribution (with $p=0.5$).
We define binary features as imaging biomarkers $x_{\mathit{pred, prog}}\in \{0,1\}$: (a) the color (green or not green) as prognostic feature and whether digits lack or contain a circle or loop (\ie $\{1,2,3,4,5,7\}$ vs. $\{0,6,8,9\}$) as the predictive feature or (b) vice versa. For intuition, a treatment might involve applying an image filter to alter the digit's appearance, while the outcome might be a digit classifier's confidence score.

\textbf{Bird species dataset (CUB-200-2011).} The dataset includes images of 200 bird species, 5,794 for testing and 5,994 for training, which we further split into training and validation data with an $80\%/20\%$ split. From the binary attributes of the birds, we select two visually distinct biomarkers  $x_{\mathit{pred, prog}}\in \{0,1\}$ with high annotator certainty: (a)~``\textit{has primary color: white}'' as prognostic and ``\textit{has bill length: longer than head}'' as the predictive feature or (b)~vice versa. To illustrate, the imaging biomarkers here might relate to the bird's observed behavior as an outcome, and habitat modification might serve as the treatment.%

\textbf{Skin lesion dataset (ISIC 2018).} The ISIC 2018 dataset contains skin lesion images with a designated training dataset of 2,594 images, which is split into a training and validation set of sizes 2,075 and 519 respectively. Final evaluations are performed on the designated validation set with 100 images.
We identify dermoscopic attributes, \ie visual skin lesion patterns, using ground truth segmentation masks and assign their presence to biomarkers. In feature set (a) the presence of globules is prognostic and the presence of a pigment network is predictive, or in (b) vice versa. Both features have been evaluated as imaging biomarkers for diagnosing melanoma~\cite{gareau2017digital,gareau2020deep} making them realistic examples of biomarkers. Unlike the features of the previous datasets, these features are based on the presence of patterns rather than localized features or color values.

\textbf{Lung cancer CT dataset (NSCLC-Radiomics).} This dataset comprises 415 3D CT volumes of pre-treatment scans from NSCLC patients and ground truth segmentation masks of the lung tumors. We crop the volumes to the largest connected tumor volume bounding box, use 332 samples for 5-fold cross-validation, and reserve 83 for testing. We define two continuous, uncorrelated radiomics features described in~\cite{zwanenburg2020image} as biomarkers, which have both been evaluated for their prognostic or predictive value before~\cite{aerts2014decoding,bortolotto2021radiomics}: (a) the shaped-based feature ``flatness'' describing the ratio between the smallest and largest principal tumor components as a prognostic feature and the first-order statistics feature ``energy'' characterizing the sum of squares of tumor intensity values as a predictive feature or (b) vice versa. The flatness feature is inverse to the actual flatness of the tumor. Values close to 0 indicate flat shapes, whereas values close to 1 indicate sphere-like shapes. Energy depends strongly on both volume and minimum pixel intensity as the minimum intensity value is added as an offset. The radiomics features were extracted from the ground truth tumor segmentation volumes with PyRadiomics~\cite{van2017computational}.

We split all datasets randomly into two equally sized subsets, a control ($T=0$) and a treatment group dataset ($T=1$), and generate group-specific outcomes $Y(T,x)$ according to Eq.~(\ref{eq:biomarker}). For each CMNIST feature, we choose the biomarker strength parameters $b_{\mathit{pred,prog}} \in \{0.0,0.1, \dots ,1.0\}$, resulting in training 121 models. For the remaining datasets, we choose the biomarker strength parameters $b_{\mathit{pred,prog}} \in \{0.0,0.2,0.4,0.6,0.8,1.0\}$, resulting in 36 different trained models.

\section{Results}
\begin{figure*}[ht!]
    \centering
    \includegraphics[width=0.96\linewidth, page = 1]{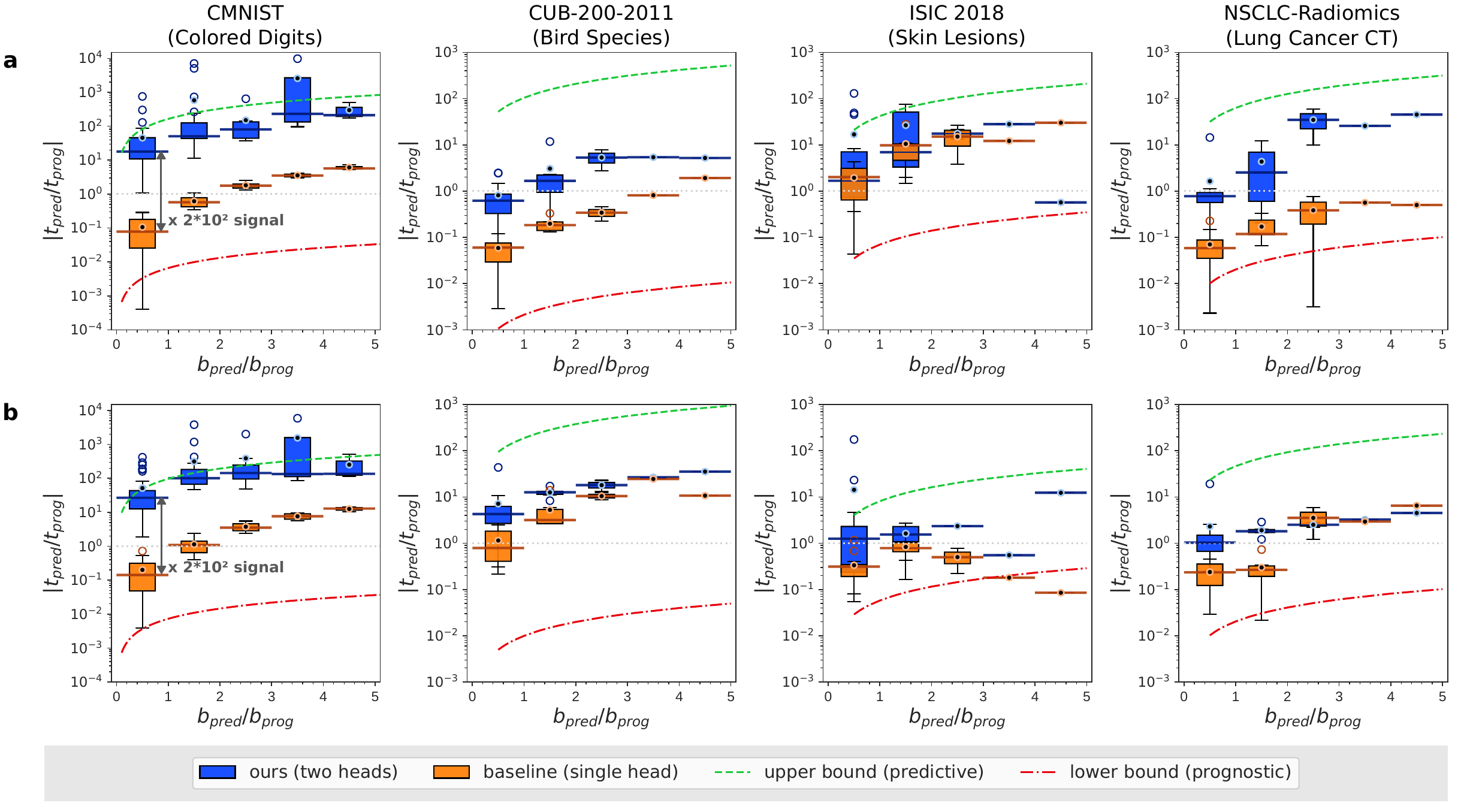}
    \caption{Model performance based on the relative predictive strength $t_{\mathit{pred}}/t_{\mathit{prog}}$ of the CATE, shown on a logarithmic scale. We compare our two-headed CATE estimator with a one-headed baseline model across different simulation parameters $b_{\mathit{pred}}/b_{\mathit{prog}}$ (\ie relative size of the predictive effect in the simulated outcomes).
    Boxplots summarize data averaged over $b_{\mathit{pred}}/b_{\mathit{prog}}$-bin widths, indicated by the horizontal error bars over the median line. Rows (\textbf{a}) and (\textbf{b}) correspond to different sets of prognostic and predictive features used for generating the data (see \cref{sec:datasets} and Fig.~\ref{fig:biomarkers}). The variance of the boxplots is affected by the differing number of samples each bin contains.
    } %
    \label{fig:results}
\end{figure*}

\subsection{Predictive strength of the estimated CATE}%

We present the results of our quantitative experimental validation protocol in Fig.~\ref{fig:results}, where the estimated relative predictive strength $|t_{\mathit{pred}}/t_{\mathit{prog}}|$ reflects its dependency on the relative size of the true predictive effect $b_{\mathit{pred}}/b_{\mathit{prog}}$ in the outcome simulation described in Fig.~\ref{fig:overview}. %
Across the four datasets, the CATE estimation model shows higher relative predictive strength $t_{\mathit{pred}}/t_{\mathit{prog}}$ with higher relative predictive biomarker signal strength $b_{\mathit{pred}}/b_{\mathit{prog}}$, often surpassing the baseline models, especially for low $b_{\mathit{pred}}/b_{\mathit{prog}}$. While the results are similar for models (a) and (b), the difference is more pronounced for the other datasets, indicating a greater influence of the type of biomarkers.

Our model performs best on CMNIST among all four datasets, with a significantly larger gap from the baseline. For example, it reaches a factor of $10^2$ for $b_{\mathit{pred}}/b_{\mathit{prog}}$ in the range of 0 to 1, and has results much closer to the upper bound than the lower bound.

While the relative predictive strength for CUB-200-2011 is lower than on CMNIST, it remains above the lower and near the upper bound. For $b_{\mathit{pred}}/b_{\mathit{prog}}$ between 0 and 1, the median $t_{\mathit{pred}}/t_{\mathit{prog}}$ differs from the baseline by factors of 10 and 5 for sets (a) and (b), respectively. The dependency on the biomarker choice is evident from the smaller gap between our model and the baseline in set (b) versus (a). 

The ISIC 2018 results show smaller absolute $t_{\mathit{pred}}/t_{\mathit{prog}}$ values, yet the relative predictive strength mean values remain above 1, except for two outliers at high $b_{\mathit{pred}}/b_{\mathit{prog}}$, based on a single sample. In set (a), the absolute $t_{\mathit{pred}}/t_{\mathit{prog}}$ values are higher and much closer to the upper bound, but exhibit greater boxplot overlaps with the baseline for low $b_{\mathit{pred}}/b_{\mathit{prog}}$ compared to set (b), where ``has globules'' is predictive. In set (b), the medians differ by a factor of 4 for relative $b_{\mathit{pred}}/b_{\mathit{prog}}$ in the range of 0 to 1. The large baseline values suggest the baseline model also strongly relies on the predictive biomarker ``has pigment networks''.

On NSCLC-Radiomics, our model demonstrates larger $t_{\mathit{pred}}/t_{\mathit{prog}}$ gaps between model and baseline, particularly for smaller $b_{\mathit{pred}}/b_{\mathit{prog}}$, with gaps decreasing slightly as $b_{\mathit{pred}}/b_{\mathit{prog}}$ increases for set (b). The performance differs between biomarker sets (a) and (b), with medians of our models and baseline differing by a factor of 13 and 4 respectively for $b_{\mathit{pred}}/b_{\mathit{prog}}$ in the range of 0 to 1.

\subsection{Interpreting predictive imaging biomarkers}\label{sec:XAIresults}
\begin{figure*}[!ht]
    \centering
    \includegraphics[trim=0cm 18cm 19cm 0cm,clip, width=0.94\linewidth]{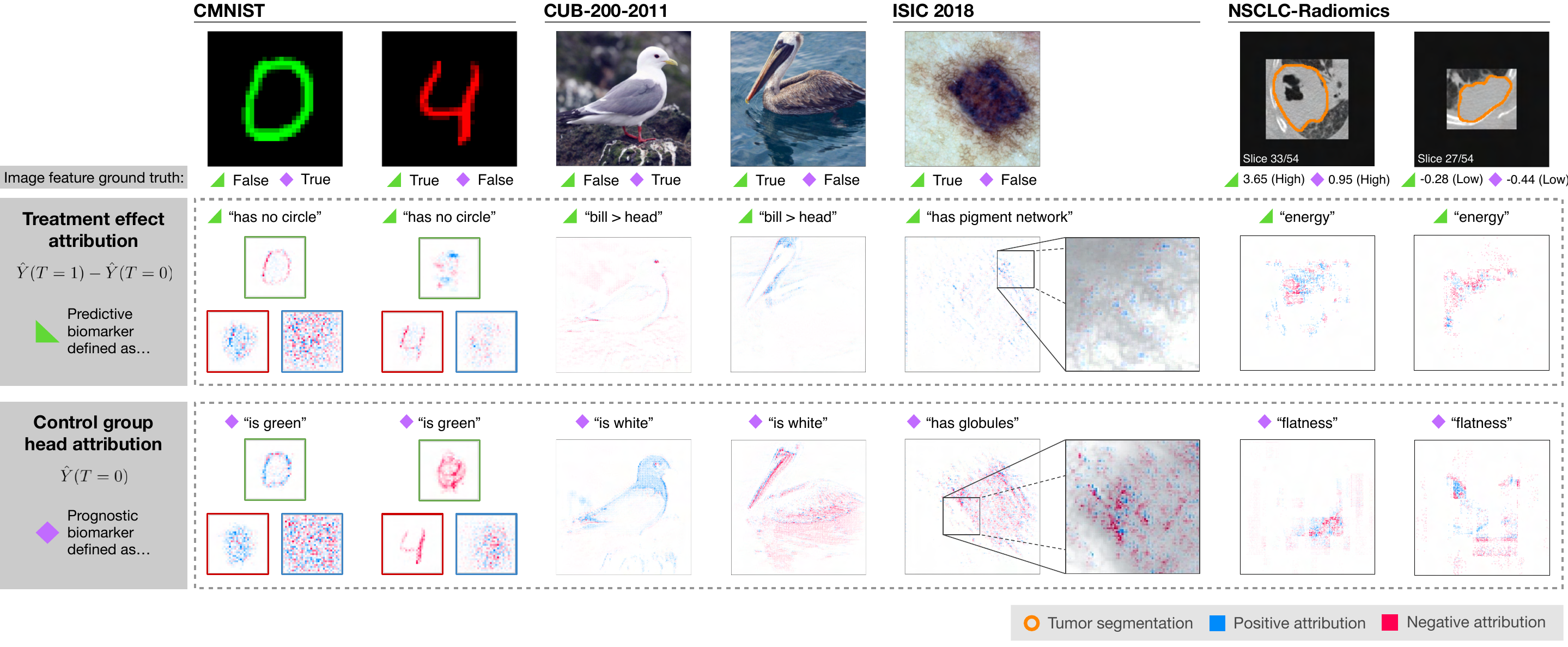}
    \caption{Attribution maps for the control group prediction head (last row) and the predicted CATE output (middle row) for different example images from each dataset (top row). For the CMNIST dataset, the attribution is shown for each RGB color channel (red: left, green: top, blue: right), as the color information is important for the biomarker prediction. An additional zoomed-in patch of the ISIC 2018 attribution map is overlaid with a grayscale version of the original image. For the NSCLC-Radiomics dataset, sagittal slices of the 3D patches are shown with orange outlines of segmented tumors. Here, results are based on models trained with $b_{\mathit{pred}},b_{\mathit{prog}} = 1.0$.}
    \label{fig:XAI}
\end{figure*}

In Fig.~\ref{fig:XAI}, we illustrate our XAI-based evaluation scheme to assess whether the image features identified by our CATE estimation model as predictive or prognostic correspond to the ground truth biomarkers.
By applying attribution methods \cite{sundararajan_axiomatic_2017,erion_improving_2020,Selvaraju2017,Springenberg2014} to our model and an input image, we generate an attribution map, indicating positive (blue) and negative (red) contributions to the prediction.
We show attribution maps of the predicted CATE, $\hat{Y}(T=1)-\hat{Y}(T=0)$, which is expected to be sensitive only to the predictive biomarker (Fig.~\ref{fig:overview}b), and the control group head, $\hat{Y}(T=0)$, which should be sensitive to the prognostic biomarker.  

For CMNIST, the attribution maps of the predicted CATE $\hat{Y}(T=1)-\hat{Y}(T=0)$ show mostly negative attribution in the green channel of the first example, which corresponds to the absence of the predictive biomarker ``has no circle'' in the input image. Similarly, the treatment effect attribution maps for the second example (red digit four) show weaker negative attribution from the digit in the red channel with some noisy positive attribution in the background. More positive attribution is observed in the green channel, indicating that the model correctly infers that the predictive biomarker ``has no circle'' is present. 
The control group head output $\hat{Y}(T=0)$ correctly identifies the prognostic biomarker, \ie ``digit is green'', in the respective color channel, which is evident from the mainly positive attribution in the green color channel in the first example and negative attribution in the red color channel for the second example. 
For both outputs' attribution maps, only noisy attribution is present for the blue channel, suggesting that the model does not use this channel for prediction.

In the first CUB-200-2011 example, where the predictive biomarker ``bill longer than head'' is absent, the attribution map for $\hat{Y}(T=1)-\hat{Y}(T=0)$ is mostly negative and focusing on the eye and outlines of the throat and breast. The attribution is not as localized as in the second example, where the predictive biomarker is present and the overall attribution is positive. Here, features of the head are primarily used for the predictions, while the main body and wings are ignored, reinforcing the importance of the bill and head region for determining the predictive biomarker.
The $\hat{Y}(T=0)$ attribution map shows overall positive attribution, especially in the white head and breast region from the first, primarily white bird. For the second bird, the attribution map is overall negative, particularly in the dark wing, main body, and pouch region. These patterns indicate that the model correctly identifies the presence or absence of the prognostic biomarker in the corresponding example. 

The ISIC 2018 image shows a pigment network surrounding a darker center. Several patterns become apparent in the attribution map overlaid with the original image. However, the allocation of positive or negative attribution provides only limited insight, possibly due to the biomarker features' complexity.
In the $\hat{Y}(T=1) - \hat{Y}(T=0)$ attribution map, positive attributions are given to the periphery surrounding the dark center where the pigment network is located. Notably, the model relies on the less pigmented gaps between the dark grid-like structures to detect the pigment network, suggesting that the gaps contain sufficient information for their detection.
The $\hat{Y}(T=0)$ attribution map reveals that the model uses the dark lesion center for control group predictions, with red and blue spots indicating the model's search for the small globule dots. 

In the first NSCLC-Radiomics example, the highest absolute values in the $\hat{Y}(T=1) - \hat{Y}(T=0)$ attribution maps are observed within the tumor area. While the attention maps show negative attributions in the darker tumor regions, positive attributions can be seen in the surrounding areas, indicating the presence of a strong predictive biomarker. This observation is consistent with the ground truth, where the energy value is comparably high, whereas mostly negative attributions are observed for the second example with a lower energy value. However, the attributions are mainly given to the areas outside the outline of the tumor, potentially due to the network's difficulty in correctly identifying the tumor boundary.
The $\hat{Y}(T=0)$ attribution maps show strong attributions mainly outside the tumor outline, which relates to the prognostic biomarker flatness. Additionally, artifacts around the border suggest that the patch shapes contribute partially to the prediction.
Further results and a more detailed qualitative XAI analysis can be found in the Supplementary section \textbf{A.2}.

\section{Discussion}

\label{sec:discussion}
The results suggest that the estimated CATE used in our quantitative evaluation approach is a reliable measure both for the predictive effect and the predictive biomarker itself under the assumption of a linear biomarker-outcome relation. The experiments also highlight how an image-based CATE estimator can be employed to identify predictive biomarkers from our simulated data while not being affected by prognostic biomarkers across various types of biomarkers and input images. This was validated by comparing the relative predictive strength $t_{\mathit{pred}}/t_{\mathit{prog}}$ to our experimental baseline as well as our experimental upper and lower bound in our proposed experiments. Even in scenarios where predictive effects are smaller than prognostic effects for $b_{\mathit{pred}}/b_{\mathit{prog}}<1$, which is often observed in real-world data, the model demonstrated the ability to identify predictive imaging biomarkers.

However, for the specific image-based CATE estimator we used, weaker performance is observed for CUB-200-2011, ISIC 2018, and NSCLC-Radiomics, particularly when $b_{\mathit{pred}}/b_{\mathit{prog}}$ is high and where $t_{\mathit{pred}}/t_{\mathit{prog}}$ is close to the baseline. This may be due to the model's lower accuracy in predicting outcomes $Y$ when facing more abstract features, along with the imbalance and distribution of image features found in the datasets, issues that could be addressed by CATE estimators designed for this purpose. %
In practical applications, where a single model is trained on data with unknown predictive effects, a quantitative evaluation would entail performing regression and $t$-tests on the parameters, as described in \cref{sec:eval}, to assess the model's ability to identify information relevant for treatment effects (\ie predictive biomarkers).

Our qualitative experimental results empirically demonstrate how an image-based CATE estimator's ability to identify predictive biomarkers can be assessed by comparing whether the treatment effect attribution maps to the selected ground truth predictive imaging biomarkers features. This is effective for both localized features based on color and shape (CMNIST, CUB-200-2011, NSCLC-Radiomics), as well as first-order statistics (NSCLC-Radiomics) or patterns (ISIC 2018).
In applications, our XAI analysis is essential for identifying and interpreting predictive and prognostic imaging biomarkers. Unlike tabular data, images lack discrete candidates for feature importance scores. Distinguishing between predictive and prognostic imaging biomarkers using attribution maps becomes challenging when located in the same image areas. The heatmap focuses on the same pixels (as with energy and flatness in the NSCLC-Radiomics example), making it difficult to discern whether an image feature that is both predictive and prognostic is present, or if two independent imaging biomarkers with distinct meanings are spatially overlapping. In such cases, other XAI methods like counterfactual explanations~\cite{goyal2019counterfactual} could quantify the effect of different properties of the same feature. Despite potential ambiguities for more abstract biomarkers, our evaluation can offer valuable insights into the features used by the model for its predictions. 

While we acknowledge the limitations of using only semi-synthetic data, due to the current unavailability of public RCT image datasets with verified predictive imaging biomarkers, we also emphasize its advantages. Semi-synthetic data enables us to demonstrate the performance of CATE estimation models in a reproducible way, as discussed in~\cite{curth2021really} and~\cite{crabbe2022benchmarking}. Our approach to predictive imaging biomarker discovery and evaluation does not rely on handcrafted features such as radiomics. Instead, we use radiomics features as biomarkers to simulate outcomes in our experiments, serving merely as a baseline for conducting performance comparisons.

\section{Conclusion}
In this paper, we introduce the task of identifying predictive imaging biomarkers and show how a candidate identified by a model can be evaluated through (1) a statistical evaluation comparing the predictive strength relative to prognostic interactions, and (2) attribution maps to support the interpretation of the identified candidate.
We outline an approach using an image-based CATE estimator to solve this task, enabling the discovery of new predictive imaging biomarkers without relying on potentially biased handcrafted features or image feature extractors. This also facilitates the detection of even abstract concepts from high-dimensional data, as demonstrated by our experiments. 
Our proposed experiments and analysis for assessing a model's qualitative and quantitative performance offer valuable insights for developing image-based CATE estimation methods tailored to specific challenges, such as adapting various network architectures for vision tasks and using CATE estimators previously applied to only tabular data.
Our evaluation provides a foundation for future research addressing different imaging modalities and problem settings. This may include addressing non-linear biomarker-outcome relations, \eg survival or time-to-event data, and mitigating confounding effects in observational data. 
Overall, we believe that applying image-based CATE estimators to discover unknown predictive biomarkers from imaging data can significantly enhance image-based treatment decision-making for personalized medicine and applications beyond.

\paragraph{Acknowledgements.}
{We acknowledge funding from the German Research Foundation (DFG) as part of the Priority Programme 2177 Radiomics: Next Generation of Biomedical Imaging (project identifier: 428223917) and Collaborative Research Center 1389 (UNITE Glioblastoma – project identifier: 404521405). PV is funded through an Else Kröner Clinician Scientist Endowed Professorship by the Else Kröner Fresenius Foundation (reference number: 2022 EKCS.17). This work was partly funded by Helmholtz Imaging (HI), a platform of the Helmholtz Incubator on Information and Data Science. We thank David Zimmerer for the feedback on the manuscript.}
{\small
\bibliographystyle{ieee_fullname}
\bibliography{egbib}
}

\end{document}